\documentclass[aps,twocolumn,superscriptaddress,prc]{revtex4}

\usepackage{graphicx}

\begin{document}

\title{Cold, dense nuclear matter in a SU(2) parity doublet model}

\author{D. Zschiesche}
\email{detlef@if.ufrj.br}

\affiliation{
Instituto de F\'{i}sica, Universidade Federal do Rio de Janeiro,
C.P. 68.528, 21941-972 Rio de Janeiro, RJ, Brazil}

\affiliation{
Institut f\"ur Theoretische Physik,
J. W. Goethe Universit\"at,
D-60438 Frankfurt am Main,
Germany}

\author{L. Tolos}
\email{l.tolos@gsi.de}

\affiliation{
Gesellschaft f\"ur Schwerionenforschung,
D-64291 Darmstadt, Germany}

\author{J\"urgen Schaffner-Bielich}
\email{schaffner@astro.uni-frankfurt.de}

\affiliation{
Institut f\"ur Theoretische Physik,
J. W. Goethe Universit\"at,
D-60438 Frankfurt am Main,
Germany}

\author{Robert D. Pisarski}
\email{pisarski@quark.phy.bnl.gov}

\affiliation{
Department of Physics,
Brookhaven National Laboratory, 
Upton, NY 11973,
U.S.A.}

\begin{abstract}
  We study dense nuclear matter and the chiral phase transition
  in a SU(2) parity doublet model at zero
  temperature.  The model is defined by adding the 
  chiral partner of the nucleon, the N', to the linear sigma model,
  treating the mass of the N' as an unknown free parameter.  
  The parity doublet model gives a reasonable
  description of the properties of cold nuclear matter, and avoids
  unphysical behaviour present in the standard SU(2) linear sigma model.
  If the N' is identified as the N'(1535),
  the parity doublet model shows a first order phase transition to a chirally
  restored phase at large densities, $\rho \approx 10 \rho_0$,
  defining the transition by the 
  degeneracy of the masses of the nucleon and the N'.
  If the mass of the N' is chosen to be 1.2 GeV, then
  the critical density of the chiral phase transition is lowered to 
  three times normal nuclear matter density, and for physical values of 
  the pion mass, the first order transition turns into a smooth crossover.
\end{abstract}

\maketitle


\section{Introduction}

The description of the properties of nuclear matter in effective models
of the nuclear interactions, such
as in Quantum HadroDynamics, has been quite successful \cite{Serot:1997}.
Early attempts to incorporate the basic symmetries of the
underlying fundamental theory of the strong interactions, QCD, 
however, have experienced severe difficulties. Lee and Wick studied the
standard SU(2) linear sigma model for nuclear matter, and showed that there
is a state where chiral symmetry is effectively restored by a
nearly vanishing nucleon mass \cite{Lee:1974ma}. It was, however,
discovered immediately by Kerman and Miller, that the solution is
unstable in the standard SU(2) linear sigma model, and that the model can
not describe nuclear matter saturation \cite{Kerman:1974yk}.

Boguta extended the linear sigma model by introducing a dynamically
generated vector meson in a chirally invariant way \cite{Boguta:1982wr}.
However, the model was unable to generate a chiral phase transition at
finite density or finite temperature \cite{Glendenning:1986} as the
chirally restored phase was mechanically unstable. With the inclusion of
an additional scalar field, the dilaton field, the unphysical
bifurcations could be avoided, although the compression modulus turned
out to be unphysically high \cite{Mishustin:1993ub}. A chiral phase
transition is present for either high temperatures or high densities. In
the temperature-density plane, however, the critical line of the chiral
phase transition was open, so that there was no chiral phase transition present
for intermediate temperatures and densities  \cite{Papazoglou:1996hf}. In
particular, when the vector meson masses were generated only by the
coupling to the sigma field, chiral symmetry restoration could not be
reached, as the vector meson mass vanishes. Also, first attempts to use a
linear sigma model to describe nuclei failed
\cite{Furnstahl:1992nu,Furnstahl:1993wx}.

Different effective interaction potentials were introduced to cure the
apparent caveats of the linear sigma model. A logarithmic potential
term was examined in \cite{Heide:1993yz}, and successfully applied to the
description of nuclear matter and nuclei. On the other hand, studies
using the standard Mexican hat potential were not able to adequately
describe nuclei. Admittedly, the logarithmic potential could not be
applied for studying the chiral limit, where the pion mass becomes
exactly massless in the Nambu-Goldstone phase.

A nonlinear realization of chiral symmetry was used in
\cite{Furnstahl:1995zb,Furnstahl:1996wv} to successfully describe 
the properties of
nuclear matter and nuclei. The essential ingredient of the
successful approach was that the scalar field was explicitly kept as a
dynamical degree of freedom to describe nuclear matter. Extensions to
the linear and nonlinear realization in chiral SU(3) symmetry were
performed in \cite{Papazoglou:1997uw,Papazoglou:1998vr} which included 
hyperon degrees of freedom and the description of hypernuclei. 
Effects from the strange
quark condensate were found to be important for arriving at a reasonable
compression modulus of nuclear matter and a reasonable description of nuclei.

In all of these investigations, it appears that the scalar sigma field and
its vacuum expectation value can not possibly serve, simultaneously,
as the chiral partner of the pion, the generator of the nucleon mass,
and the mediator of scalar attraction for nucleons. This
problem emerged to be particular evident when extending the model to
strange baryons, in trying to describe both the masses and the
potentials for hyperons. Inclusion of an additional scalar field,
the dilaton field, does not remedy the situation
\cite{Papazoglou:1998vr}.

An alternative way of looking at the role of the sigma field for the
generation of the hadron masses is established by two seemingly
different chiral approaches: a hidden local symmetry for the vector
mesons, and a parity doublet model for nucleons.

A hidden local symmetry is a type of gauged linear sigma model:
the overall mass scale of the vector and axial vector fields is
fixed by a new constant, while their {\em splitting} is due to a
nonvanishing vacuum
expectation value for the sigma field. This model automatically exhibits
vector meson dominance and gives a good description of
the vacuum properties of the vector and axial vector mesons
\cite{Ko:1994en}. The model was extended by the Minnesota group to
describe low-energy pion-nucleon scattering, the properties of nuclei, and
nuclear matter at finite temperature and density
\cite{Carter:1995zi,Carter:1996rf,Carter:1997fn}. Note that the chiral
symmetry demands only that the spectral functions of the vector and
axial vector mesons are degenerate in the chirally restored phase, and
not that the vector and axial vector meson masses drop to zero (see
e.g.\ \cite{Pisarski:1995xu}).

In a parity doublet model for nucleons, the chiral partner of the
nucleon, the N', is added to the linear sigma model.
This possibility was raised by Sakurai, with the
first realistic model 
given by DeTar and Kunihiro \cite{DeTar:1988kn}.
In a certain
assignment, to be presented below, the sigma field is responsible for
causing the {\em mass splitting} between the nucleon and the N', while the
overall mass scale of the nucleons 
is given by a new parameter, which couples in a chirally
invariant fashion \cite{Jido:1998av}.  The N', which is a 
negative parity state, is usually taken to have the mass of the
N'(1535).  
The parity doublet model was extended to successfully
describe resonances \cite{Jido:1999hd}, the baryon octet
\cite{Nemoto:1998um}, and medium effects of
the N' at finite densities 
\cite{Kim:1998up,Jido:2002yb,Nagahiro:2003iv}. 
For this model, nuclear matter and the
chiral phase transition in cold, dense systems 
were studied so far only in Ref.\ \cite{Hatsuda:1988mv}.  An
extended scalar interaction potential with a logarithmic term was used
to find stable solutions for nuclear and neutron matter, and 
to study the chiral phase transition at high densities.

Generically, the sigma field is now 
only responsible for mass splittings: 
this alleviates many of the problems 
present in the standard linear sigma model, and is the focus of the
present work.  We will study the
properties of the parity doublet model for describing nuclear matter and
for the chiral phase transition. We demonstrate that the instabilities of
the standard linear sigma model, with a Mexican hat potential, are
avoided when the N' is introduced as the chiral partner of the nucleon.
In addition, we study the consequences of a light N' on the
properties of nuclear matter and the chiral phase transition. We note
that the masses of the known chiral partners of the pseudoscalar and
scalar mesons are split by about 
300 to 400 MeV.  Further, all of the scalar mesons are 
extremely broad, so they are difficult to identify experimentally.
According the Particle Data Group \cite{PDBook},
the sigma meson mass lies somewhere between 400 and
1200 MeV. A successful description
of the density distribution of nuclei demands that the mass of the sigma
meson must be close to 500 MeV \cite{Furnstahl:1995zb}. The mass
splitting between the $\eta-a_0$ and the $K-\kappa(800)$, as well as that
for strange D-mesons, are all in similar ranges, 300 to
400 MeV (see \cite{PDBook}). The presumptive chiral partner of the
nucleon, the N'(1535), is located nearly 600 MeV above the nucleon mass.
The N'(1535) is not very broad, contrary to the sigma or $\kappa$
mesons, and decays by a large fraction to a nucleon and an $\eta$. The
phase space for the $\eta N$ decay is heavily suppressed compared to the
$\pi N$ decay, so that the coupling strength of the N'(1535) to the
nucleon and the $\eta$ must be unnaturally large. The chiral partner of
the nucleon should have a strong coupling to the pion, which is
seemingly absent for the N'(1535).  Therefore, we suggest that the true
chiral partner of the nucleon might be closer in mass to the nucleon
with a similar width as for the sigma meson, so that it escaped
experimental detection so far. 
It turns out that
a smaller mass for the N' reduces the critical density 
for the chiral phase considerably.

This paper is organized as follows: first we introduce the chiral SU(2)
parity model and fix its parameters. For comparison, we study the
pressure of cold nuclear matter in the 
standard linear sigma model, and discuss the apparently unphysical
nature of its stable solutions.
We then show how a parity doublet model gives a reasonable and
stable description for the properties of nuclear matter. We 
then extend the model to high density, and study the chiral phase
transition in cold nuclear matter. 


\section{The SU(2) parity model}

There are two ways of assigning chiral transformations for
parity doubled nucleons.
In the naive assignment, the two
nucleons belong to different multiplets, while in the 
mirror assignment they 
belong to the same multiplet, and so are true chiral partners 
\cite{Jido:1998av,Nemoto:1998um,Kim:1998up}.  We adopt the latter.

In the mirror model,
under $SU_L(2) \times SU(2)_R$ transformations $L$ and $R$,
the two nucleon
fields $\psi_1$ and $\psi_2$ transform as 
\begin{eqnarray}
\psi_{1R} \longrightarrow R \psi_{1R} \  & , \hspace{1cm} &   \psi_{1L}
\longrightarrow L \psi_{1L} \ , \label{mirdef1} \\
\psi_{2R} \longrightarrow L \psi_{2R} \  & , \hspace{1cm} &   \psi_{2L}
\longrightarrow R \psi_{2L} \ . \label{mirdef2}
\end{eqnarray}
This allows for a chirally invariant mass, $m_0$:
\begin{eqnarray}
&&m_{0}( \bar{\psi}_2 \gamma_{5} \psi_1 - \bar{\psi}_1
      \gamma_{5} \psi_2 ) = \nonumber \\
&& m_0 (\bar{\psi}_{2L} \psi_{1R} -
        \bar{\psi}_{2R} \psi_{1L} - \bar{\psi}_{1L} \psi_{2R} +
        \bar{\psi}_{1R} \psi_{2L}) \ . \label{chinvmass}
\end{eqnarray}
The chiral Lagrangian in the  mirror model is
\begin{eqnarray}
{\cal L} &=& \bar{\psi}_1 i {\partial\!\!\!/} \psi_1 
+ \bar{\psi}_2 i {\partial\!\!\!/} \psi_2 \nonumber\\
&+& m_0 \left(\bar{\psi}_2 \gamma_5 \psi_1 - \bar{\psi}_1 \gamma_5
  \psi_2\right)\nonumber\\ 
&+& a \bar{\psi}_1 \left(\sigma + i \gamma_5 \vec{\tau}
  \cdot\vec{\pi}\right) \psi_1
+ b \bar{\psi}_2 \left(\sigma - i \gamma_5 \vec{\tau}
  \cdot\vec{\pi}\right) \psi_2\nonumber\\ 
&-& g_{\omega} \bar{\psi}_1 \gamma_{\mu} \omega^{\mu} \psi_1
- g_{\omega} \bar{\psi}_2 \gamma_{\mu} \omega^{\mu} \psi_2 \nonumber \\
&+& {\cal L}_M \ ,
\label{lagrangian}
\end{eqnarray}
where $a$, $b$ and $g_{\omega}$ are the coupling constants of the mesons
fields  ($\sigma$, $\pi$ and $\omega$) to the baryonic fields $\psi_1$
and $\psi_2$.  Note that we assume the same vector coupling strength for 
both parity partners. The 
mesonic Lagrangian ${\cal L}_M$ contains the kinetic terms 
of the different meson 
species, and potentials for the scalar and vector fields.
The potential for the spin zero fields is the same as in the ordinary
SU(2) linear sigma model.  
Kinetic and potential terms are added for an isoscalar vector
meson, $\omega$, 
as in the $\sigma$-$\omega$ model of nuclear matter \cite{Walecka:1974qa}:
\begin{eqnarray}
{\cal L}_M&=&\frac{1}{2} \partial_{\mu} \sigma^{\mu} \partial^{\mu}
\sigma_{\mu}  
+ \frac{1}{2} \partial_{\mu} \vec{\pi}^{\mu} \partial^{\mu}
\vec{\pi}_{\mu}  
- \frac{1}{4} F_{\mu \nu} F^{\mu \nu} \nonumber \\
&+& \frac 12 m_\omega^2 \omega_{\mu} \omega^{\mu} + g_4^4 (\omega_{\mu}
\omega^{\mu})^2 \nonumber\\ 
&+& \frac 12 \bar{\mu}\,^2 (\sigma^2+\vec{\pi}^2) - \frac \lambda 4
(\sigma^2+\vec{\pi}^2)^2  \nonumber \\ 
&+& \epsilon \sigma \ , 
\end{eqnarray} 
where $F_{\mu \nu}=\partial_{\mu}
\omega_{\nu}-\partial_{\nu}\omega_{\mu}$ represents 
the field strength tensor of the vector field.  
As usual, the parameters $\lambda$, $\bar{\mu}$ and $\epsilon$ 
can be related to the sigma and pion masses, and the pion decay 
constant, in vacuum:
\begin{eqnarray}
\lambda&=&\frac{m_{\sigma}^2-m_{\pi}^2}{2 \, \sigma_0} \nonumber \ ,\\
\bar{\mu}\,^2&=&\frac{m_{\sigma}^2-3 m_{\pi}^2}{2} \nonumber \ ,\\
\epsilon&=&m_{\pi}^2 f_{\pi} \ , 
\end{eqnarray} 
with $m_{\pi}=138$ MeV, $f_{\pi}=93$ MeV and $\sigma_0=f_{\pi}$
the  vacuum expectation value of 
the sigma field. Since the mass of the $\sigma$ meson 
in the vacuum can not be fixed precisely by experiment, we will treat 
it as a free parameter. 
The vacuum mass of the $\omega$ field is $m_{\omega}=783$ MeV while the
$g_4$ term for the $\omega$ field also represents a fit-parameter, 
with finite values of this parameter causing a softening of the 
equation of state.

To investigate the properties of dense nuclear matter and the chiral
phase transition at zero temperature, we adopt a mean-field approximation
\cite{Serot:1984ey}. The fluctuations around constant vacuum expectation
values of the mesonic field operators are neglected, while the nucleons
are treated as quantum-mechanical one-particle operators.  Only the
time-like component of the vector meson $\langle \omega \rangle \equiv
\omega_0 $ survives, assuming  homogeneous and isotropic infinite
nuclear matter.  Additionally, parity conservation demands $\langle \pi
\rangle$$=$0.

The mass eigenstates for the 
parity doubled nucleons, the $N^+$ and $N^-$, are determined
by diagonalizing the mass matrix, Eq.~(\ref{chinvmass}), 
for $\psi_1$ and $\psi_2$:
\begin{eqnarray}
\left(\begin{array}{c} N^+ \\ N^- \end{array}\right)=
\frac{1}{\sqrt{2 \, {\rm cosh \delta}}}
\left(\begin{array}{cc} e^{\delta/2} & \gamma_5 \,  e^{-\delta/2}\\
\gamma_5 \, e^{-\delta/2} & -e^{\delta/2} \end{array}\right)
\left(\begin{array}{c} \psi_1 \\ \psi_2 \end{array} \right)
\label{n-psi}
\end{eqnarray}
where ${\rm sinh \delta}=-(a+b)\sigma/2 m_0$. In the basis of 
Eq.~(\ref{n-psi}) the masses of $N^+$ and $N^-$ are given by
\begin{equation}
m_i=m_{N\pm}= \frac 12 \left( \sqrt{(a+b)^2 \sigma^2 + 4 m_0^2} \mp
  (a-b) \sigma \right)\ .  
\end{equation}
If chiral symmetry is completely restored, i.e. $\sigma=0$,
the two nucleonic parity states become degenerate in mass with 
$m_{N^+}=m_{N^-}=m_0$. Thus, if the value of $m_0$ is large, 
the nucleon masses are primarily
generated by the explicit mass term, while 
the spontaneous breaking only generates
the mass splitting ~\cite{Jido:1998av}.
 In contrast, in the naive assignment, as well as in 
the standard linear sigma model (and also the mirror model
model with $m_0=0$), spontaneous symmetry breaking 
generates the nucleonic masses.

The grand canonical partition function is 
\begin{eqnarray} 
\frac{\Omega}{V}={\cal V_{\rm M}}+\sum_i \frac{\gamma_i}{(2 \pi)^3}
\int_{0}^{k_{F_i}} \, 
d^3k \, (E_i^*(k)-\mu_i^*) \ , 
\end{eqnarray} 
where $i \in \{N^+,N^-\}$ denotes the nucleon type, 
$\gamma_i$ is the fermionic degeneracy, 
$E_{i}^* (k) = \sqrt{k^2+{m_i}^2}$ the energy, 
and $\mu_i^*=\mu_i-g_{\omega} \omega_0=\sqrt{k_F^2+m_i^2}$ 
the corresponding effective chemical potential.
The single
particle energy of each parity partner $i$ is given by
$E_i(k)=E_i^*(k)+g_{\omega} \omega_0$.

The mean meson fields $\bar{\sigma}$ and $\bar{\omega}$ are determined by extremizing the
grand canonical potential $\Omega/V$: 
\begin{eqnarray}
\left . \frac{\partial(\Omega/V)}{\partial \sigma} \right |_{\bar{\sigma},\bar{\omega}} 
&=&-\bar{\mu}\,^2 \bar{\sigma} +
\lambda \bar{\sigma}^3 - \epsilon + \sum_i \rho_i^*(\bar{\sigma},\bar{\omega}) 
\left . \frac{\partial m_i}{\partial  \sigma} \right |_{\bar{\sigma}} =0 \ ,\nonumber \\ 
\left . \frac{\partial(\Omega/V)}{\partial \omega_0} \right |_{\bar{\sigma},\bar{\omega}} &=&-m_\omega^2
\bar{\omega} -4 g_4^4 \bar{\omega}^3+g_{\omega} \sum_i \rho_i (\bar{\sigma},\bar{\omega})=0 \ . 
\label{eqsmotion} 
\end{eqnarray} 
The scalar density $\rho_i^*$ and the baryon density $\rho_i$ for each
chiral partner are given by the usual expressions 
\begin{eqnarray}
  \rho_i^*&=&\gamma_i \int_0^{k_{F_i}} \frac{d^3k}{(2 \pi)^3}
  \,\frac{m_i}{E^*_i}\nonumber \\ 
  &=&\frac{\gamma_i m_i}{4 \pi^2}
  \left[k_{F_i} E_{F_i}^*-m_i^2 {\rm ln} \left(
      \frac{k_{F_i}+E_{F_i}^*}{m_i} \right) \right] \ , \nonumber \\ 
  \rho_i&=&\gamma_i \int_0^{k_{F_i}} \frac{d^3k}{(2
    \pi)^3}=\frac{\gamma_i k^3_{F_i}}{6 \pi^2} \ .
\label{densities}
\end{eqnarray}

The basic nuclear matter saturation properties 
we impose can be formulated in the following way. 
The stable minimum of the grand canonical potential for 
$\mu_B=923$ $\rm{MeV}$ has to meet two conditions:
\begin{eqnarray} 
\label{nucmatprop}
E/A (\mu_B=923 \rm{ MeV}) - m_N &=& -16 \rm{ MeV} \nonumber \\ \,
\rho_0 (\mu_B=923 \rm{ MeV})  &=& 0.16 \mbox{ fm}^{-3} \ .  
\end{eqnarray} 
Altogether we have four parameters to be related to 
nuclear matter properties: $m_0$, $g_{\omega}$,
$m_\sigma$ and $g_4$. The parameters $a,b$ are related 
to the vacuum masses of the parity partners. 
The mass of the positive parity state will always be the nucleon mass 
$m_{N^+}=939$~MeV. The negative parity state will  be a free parameter. Here 
we will consider two cases: 
$m_{N^-}=1.5$~GeV, which corresponds to the conventional 
choice of N'(1535) as the parity partner and, 
as an alternative, $m_{N^-}=1.2$~GeV.
We will investigate 
if a pair $(g_{\omega},m_\sigma)$ exists which 
fulfills Eq.~(\ref{nucmatprop}) for a given choice of $m_0$, 
$g_4$ and $m_{N^-}$. 
Then, by varying  $m_0$ and $g_4$, we check 
how the fit parameters or 
observables like the incompressibility 
\begin{eqnarray} 
K=9 \rho_0^2 \left.  \frac{\partial^2 (E/A)}{\partial \rho^2}
\right|_{\rho=\rho_0}
=9 \left. \frac{P}{\rho}\right|_{\rho=\rho_0}= 9\rho_0
\left. \frac{\mu_B}{\rho}\right|_{\rho=\rho_0}    
\end{eqnarray} 
change.


\section{Nuclear matter in linear sigma models}

\subsection{Nuclear matter in the standard linear sigma model}

\begin{figure}[h]
\vspace{-1.0cm}
\includegraphics[width=9cm]{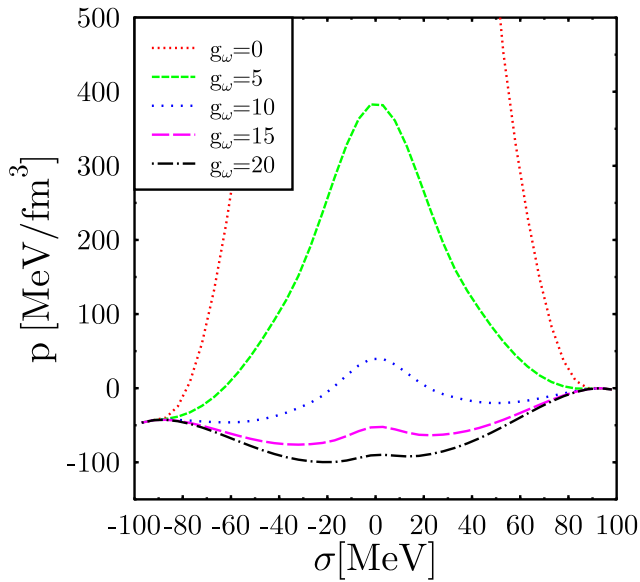}
\vspace{-0.6cm}
\caption{\label{press_m0}
Pressure at $\mu_B=923$ MeV and $m_0=0$, versus
the chiral condensate $\sigma$,
for different values of the vector coupling $g_\omega$.}
\end{figure}

As was already shown by Kerman and Miller,
it is not possible to 
reproduce stable nuclear matter properties in the standard linear sigma model 
by 'just adding' vector mesons \cite{Kerman:1974yk}. The same also holds
for the parity model with $m_0=0$.
As already mentioned above,
for a successful description of 
nuclear matter, the minimum requirements are:  
A binding energy of $16$~MeV  
and a nuclear matter density of approximately $0.16$~fm$^{-3}$
at $\mu_B=923$~MeV. Furthermore, one would expect this 
state to be different from the vacuum or the chirally restored 
phase.
\begin{figure}[h]
\vspace{-1.0cm}
\includegraphics[width=9cm]{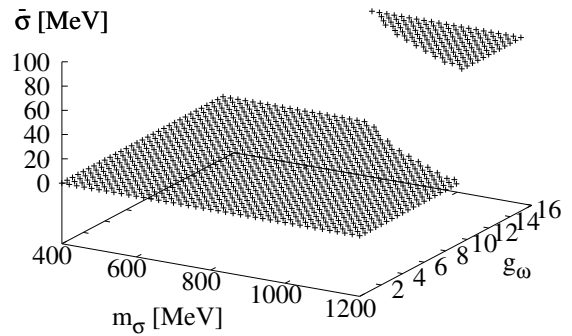}
\vspace{-0.6cm}
\caption{\label{sigma_vs_3d_m0}
Expectation value $\bar{\sigma}$ of the 
chiral condensate in the stable 
phase as a function of 
$m_\sigma$ and $g_\omega$ for $m_0=0$; $g_4=0$ and $m_{N^-}=1.5$ GeV.
There are only two possible values of the 
condensate: $\bar{\sigma}=f_\pi$ and $\bar{\sigma} \approx 0$.}
\end{figure}
But in the standard linear $\sigma-\omega$ model,
as well as in the parity model with $m_0=0$, 
for $\mu_B=923$~MeV only two possible phases exist: 
the vacuum state, characterized by $\bar{\sigma}=f_\pi$ and $\rho_B=0$,
and the chirally restored phase with $\bar{\sigma}\approx 0$. 
Which of these phases is stable, depends on the choice of 
parameters.  Neither of these phases can represent saturated nuclear matter.   

Figure \ref{press_m0} shows the pressure,
as a function of the condensate in the parity model, at $\mu_B=923$~MeV
and $m_0=0$, for different values of the vector coupling.
The remaining parameters are chosen as
$m_\sigma=1$~GeV, $g_4=0$ and $m_{N^-}=1.5$~GeV. 
For small values of $g_\omega$, 
the pressure is maximal at 
$\sigma \approx 0$, i.e. the chirally restored phase is stable.
For larger $g_\omega$ the pressure of this phase is reduced and finally 
falls below zero. Then the vacuum has the greatest pressure, and thus 
represents the stable phase. For any $g_\omega$, a stable, intermediate
phase, which could represent ordinary nuclear matter, does not exist.
This 
situation does not change if the value of the sigma
mass is varied. Figure \ref{sigma_vs_3d_m0} shows the 
mean sigma value corresponding to the stable state (maximum pressure) 
of the system as a function of $g_\omega$ and $m_\sigma$. All 
other parameters remain the same as before.
Still the stable state is either the vacuum or the phase with 
$\bar{\sigma} \approx 0$.
Note that changing the values of $m_{N^-}$ or $g_4$ does 
not alter these findings.

\subsection{Nuclear matter with the parity doublet model}

\begin{figure}[h] 
\vspace{-1.0cm}
\includegraphics[width=9cm]{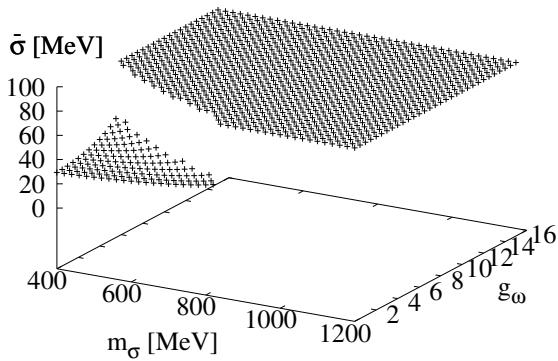}
\vspace{-0.6cm}
\caption{\label{sigma_vs_msigma_m800}
Expectation value $\bar{\sigma}$ of the chiral condensate 
in the stable phase 
as a function of 
$m_\sigma$ and $g_\omega$ for $m_0=800$.  Note the
appearance of an intermediate phase, with $\bar{\sigma} \sim 30$ MeV.
}
\end{figure}

The situation changes if we allow for finite values of the 
mass term $m_0$. As shown in Figure \ref{sigma_vs_msigma_m800}, 
when $m_0=800$~MeV, it is possible to find values of $g_\omega$
and $m_\sigma$ such that there is an intermediate phase
which might represent nuclear matter.  In this intermediate
phase, $\bar{\sigma} \sim 30$~MeV.

\begin{figure}
\vspace{-1.0cm}
\includegraphics[width=9cm]{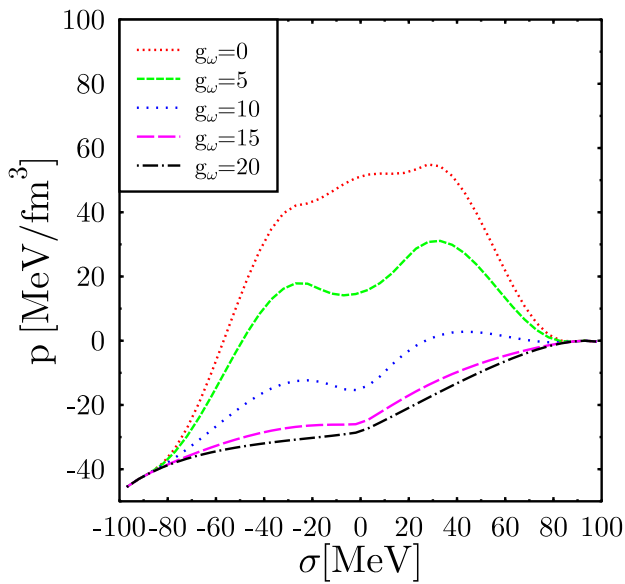}
\vspace{-0.6cm}
\caption{\label{press_m800}
Pressure at $\mu_B=923$ MeV and $m_0=800$~MeV,   
versus the chiral condensate $\sigma$,
for different values of the vector coupling $g_\omega$.
}
\end{figure}

Figure \ref{press_m800} shows the pressure as a function 
of the $\sigma$-field for
$m_0=800$~MeV, $m_\sigma = 400$~MeV, $g_4=0$ 
and $m_{N^-}=1.5$~GeV.
For small values of $g_\omega$ and $m_\sigma$, the stable phase has
$\bar{\sigma} \approx 30$~MeV; for large values of these parameters,
again the vacuum is the stable state.

\begin{figure}[h]
\vspace{-1.0cm}
\includegraphics[width=9cm]{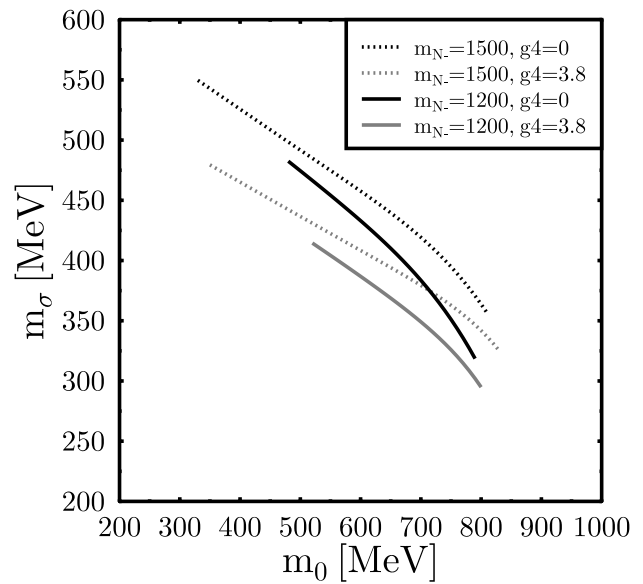}
\vspace{-0.6cm}
\caption{\label{msigma_vs_m0}
The fitted sigma vacuum mass $m_\sigma$ versus the mass parameter $m_0$. 
To describe nuclear matter, $m_0$ must be 
in the range $~300-800$~MeV for $m_{N^-}=1.5$~GeV and
$~500-800$~MeV for $m_{N^-}=1.2$~GeV.}
\end{figure}

\begin{figure}[h]
\vspace{-1.0cm}
\includegraphics[width=9cm]{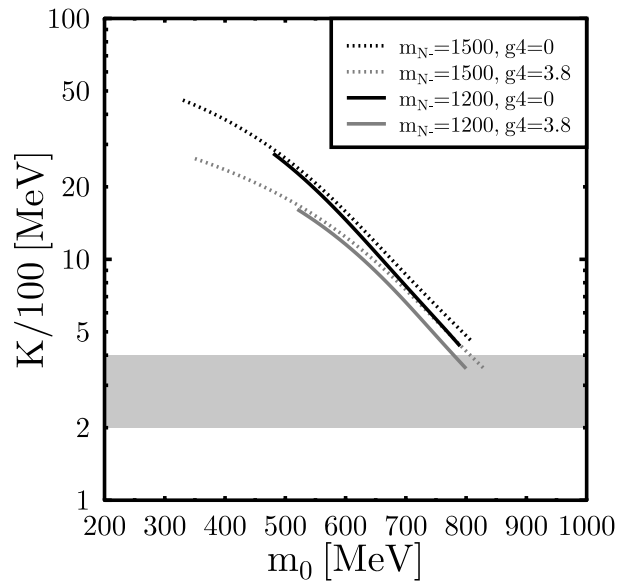}
\vspace{-0.6cm}
\caption{\label{comp_vs_m0}
Nuclear incompressibility K vs $m_0$.
The grey band shows the range of 
``allowed'' values as suggested experimentally. 
For small $m_0$ values the 
incompressibility is high. With $g_4=0$ the minimum 
is at K$\approx 450$~MeV. For $g4=3.8$, it decreases to  
K$\approx 350$~ MeV.}
\end{figure}

It turns out that, for these intermediate phases, which 
exist for a wide range of finite $m_0$ values,  
a reproduction of nuclear matter properties is possible. 
Figure 
\ref{msigma_vs_m0}  shows the $m_\sigma$
values as resulting from such fits
as a function 
of $m_0$ for $m_{N^-}=1200,1500$~MeV and $g_4=0,3.8$.
Each choice of $m_{N^-}$ and $g_4$ corresponds 
a minimum value of $m_0$, for which a nuclear matter fit exists. 
These minimum $m_0$ values lie between 300 and 500 MeV, 
with a higher $N^-$ mass 
allowing for smaller values. The maximum possible values for 
$m_0$ are in the range of 800~MeV for all cases considered.
The corresponding sigma vacuum masses are in the 
range of 300 to 550 MeV. 
The larger the sigma vacuum mass is,   
the larger the N' mass is chosen 
and it decreases with increasing $g_4$ coupling.
For high $m_0$ values the sigma mass turns out to be rather low.
The value of the corresponding nuclear incompressibility
is depicted in Fig. \ref{comp_vs_m0} as a function 
of $m_0$. Only for large values of $m_0$ it lies in 
a reasonable range. 
The same holds for finite $g_4$ values, 
although finite values lead to lower incompressibility. 
Thus, reasonable values for the incompressibility are obtained for high values of the mass parameter $m_0$ and that leads to a small vacuum sigma mass.
The situation could change if the model is extended to 
SU(3), as shown in 
\cite{Papazoglou:1997uw,Papazoglou:1998vr}.   

\begin{figure}[h]
\vspace{-1.0cm}
\includegraphics[width=9cm]{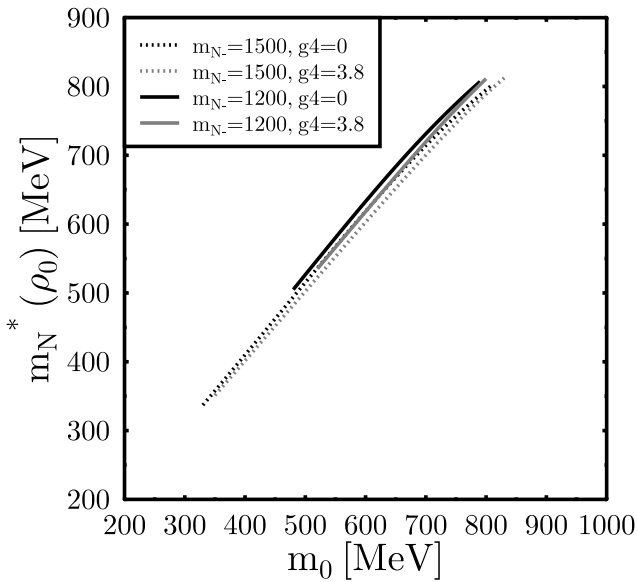}
\vspace{-0.6cm}
\caption{\label{emn_vs_m0}
The effective nucleon mass at saturation,
$m_N^\ast (\rho_0)$, versus $m_0$.}
\end{figure}

For all cases studied, Fig. \ref{emn_vs_m0} shows that
if a fit to nuclear matter is possible,
at the saturation point of nuclear matter,
the effective nucleon mass is almost identical to the value of
$m_0$.  Thus, high values of $m_0$ correspond to 
small incompressibilities, and large effective nucleon mass.
This could be a problem for the spin-orbit splitting;
as shown by Furnstahl et al \cite{Furnstahl:1997tk}, though,
such a problem can be solved by adjusting the corresponding
tensor coupling.

The results for the fits which give low incompressibilities and which 
are used in the next section to investigate dense 
matter are shown in Table \ref{fitvalues}.
\begin{table}[h]
\begin{center}
\begin{tabular}{c|c|c|c|c}
     & P1 & P2 & P3 & P4 \\
\hline
$m_{N^-}$ [MeV]  & 1200 & 1200 & 1500 & 1500 \\
$g_4$            & 0    & 3.8  & 0    & 3.8  \\
$m_0$ [MeV]      & 790  & 790  & 790  & 790  \\
\hline
$m_\sigma$ [MeV] & 318.56 & 302.01 & 370.63 & 346.59 \\
$g_{N\omega}$    & 6.08   & 6.77   & 6.79   & 7.75\\
a                & 9.16   & 9.16   & 13.00  & 13.00\\
b                & 6.35   & 6.35   & 6.97   & 6.97 \\
$\bar{\mu}$ [MeV]& 147.50 & 128.93 & 199.26 &176.29\\
$\lambda$        & 4.75   & 4.16   & 6.82   & 5.82\\
\hline
$m_{N^+}(\rho_0)/m_{N^+}$  & 0.86 & 0.86 & 0.84 &0.83\\
$m_{N^-}(\rho_0)/m_{N^-}$  & 0.79 & 0.78 & 0.73 & 0.72\\
$K$ [MeV]               & 436.41 & 374.75 & 510.57 & 440.51\\
\end{tabular}
\caption{\label{fitvalues} \textrm{
Fit parameter and nuclear matter properties for the 
four fits mainly used. For all parameter sets: $E/A(\rho_0)-m_N= -16$~MeV,   
$\rho_0 = 0.16 \mbox{ fm}^{-3}$} and the vacuum nucleon mass $m_N=939$~MeV.}
\end{center}
\end{table}

\section{The chiral phase transition}

We now investigate dense hadronic matter within the parity doublet model.
We want to use the same $m_0$ value for all cases, with the best possible
values for the incompressibility.  Thus 
we chosse 
$m_0=$790~MeV, which is  
the maximum value allowing for   
a fit to nuclear matter in all of the cases considered:
$g_4=0$ and $=3.8$, and $m_{N^-}=1.2$ and $=1.5$~GeV.
\begin{figure}[h]
\vspace{-0.5cm}
\includegraphics[width=9cm]{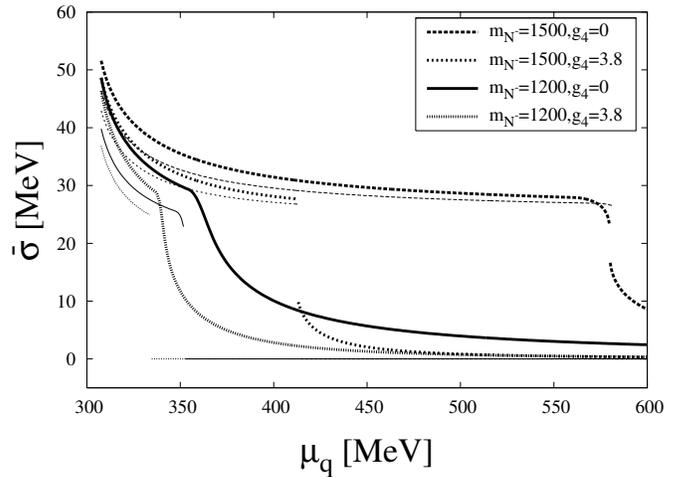}
\vspace{-0.6cm}
\caption{\label{sigma_vs_mu}
Expectation value $\bar{\sigma}$ of the chiral condensate vs quark chemical potential 
$\mu_q = 1/3 \mu_B$. 
Thick lines correspond to 
physical pion masses, thin lines to the chiral limit.
Since only the field value in the stable phase (no mixed phase) are shown, 
a discontiunity appears for a first order phase transition. }
\end{figure}
In Figure \ref{sigma_vs_mu} we show the expectation value 
of the chiral condensate as a function 
of the chemical potential. At $\mu_q = \mu_B/3 \approx 308$~MeV the condensate 
jumps from its vacuum value down to values of  around $40$~MeV, 
depending on the parameter set. This is the liquid gas phase transition, 
which is present in all cases. 
As the chemical potential is increased,
the scalar condensate decreases, until a chiral phase transition occurs.

In the chiral limit of zero pion mass, 
drawn as thin lines in Figure \ref{sigma_vs_mu},
the chiral transition is always of first
order, as the condensate jumps from $\approx 30$~MeV down to zero.

The case of a 
physical pion mass is drawn as thick lines in Figure \ref{sigma_vs_mu}.
The order of the transition, and 
the value of the critical chemical potential,
depend on the values of the $m_{N^-}$ and $g_4$,
although it is always in the range $330$~MeV$< \mu_q^c < 600$~MeV. 

When $m_{N^-}=1500$~MeV, a first order chiral transition occurs, with 
the critical chemical potential $\mu_q^c \approx 575$~MeV, for $g_4=0$ 
and $\mu_q^c\approx 410$~MeV for $g_4=3.8$.
For $m_{N^-}=1200$~MeV, the chiral ``transition'' 
becomes a smooth crossover. 
In this case we define the ``critical'' chemical potential 
as the value $\mu_q^c$ at which the change in the 
sigma field with chemical potential is the largest.
The resulting values for the  choice $m_{N^-}=1200$~MeV
are:  $\mu_q^c \approx 360$~MeV for $g_4=0$ and 
$\mu_q^c \approx 340$~MeV for $g_4=3.8$, i.e. considerably
smaller than in the case with  $m_{N^-}=1500$~MeV.
The critical chemical potentials do not change considerably 
comparing physical and vanishing pion mass for given 
$m_{N^-}$ and $g_4$.

For physical pion mass and 
$g_4=3.8$, the critical $N^-$-mass, i.e. the mass where the 
first order transition turns into a smooth transition, turns out to be 
$\approx 1490$ MeV. This explains why the 
the first order phase transition for  $m_{N^-}=1500$~MeV is rather weak.
In contrast, for $g_4=0$, the critical $N^-$-mass is found to be approximately 
$1370$~MeV, i.e. considerably away from the value of $1.5$ GeV.
Thus, as shown in Fig. \ref{sigma_vs_mu}, the discontinuity in the 
$\bar{\sigma}$ and thus the transition is 
considerably stronger.
\begin{figure}[h]
\vspace{0cm}
\includegraphics[width=9cm]{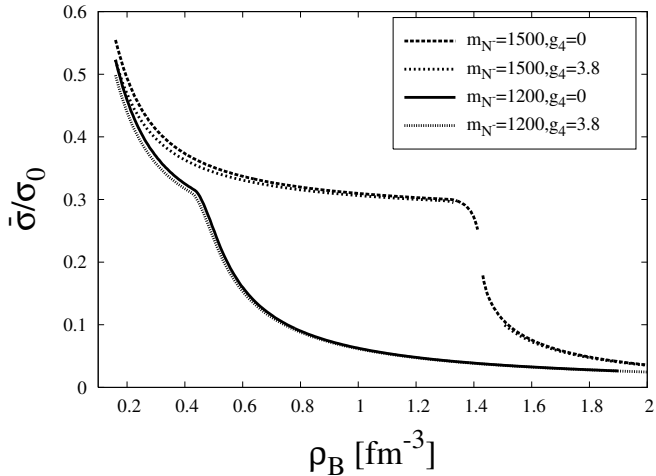}
\vspace{-0.5cm}
\caption{\label{sigma_vs_rho}
Scaled expectation value of the chiral condensate 
$\bar{\sigma}/\sigma_0$ vs the baryon density $\rho_B$.
The behaviour is nearly independent of $g_4$.
For $m_{N^-}=1.5$~GeV the transition is first order and 
happens around $8-9 \rho_0$, while
for $m_{N^-}=1.2$~GeV it is a continuous transition 
with the peak in the derivative 
$\partial \bar{\sigma}/\partial \mu$ appearing at  
$\rho \approx 3 \rho_0$. }
\end{figure}

As can be seen in Figure \ref{sigma_vs_rho}, although 
the critical chemical potentials vary considerably with the change of 
the quartic coupling, the corresponding ``critical density''
does not. This is because there is no strong coupling between 
scalar and vector field in our model.
In contrast, the critical density for the chiral transition
very strongly depends on the vacuum mass of the parity partner. 
It is reduced by a factor of three when changing the $N^-$-mass 
from 1.5 to 1.2 GeV.

\begin{figure}[h]
\includegraphics[width=9cm]{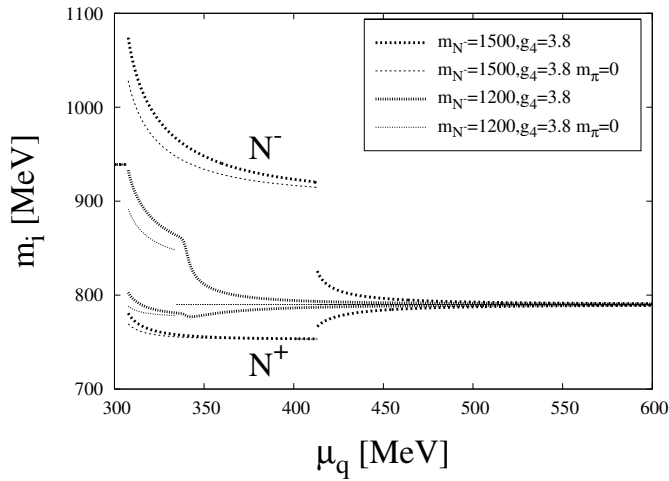}
\caption{\label{emn_vs_mu}
Effective masses of the parity partners 
versus the chemical potential.
Thick lines correspond to 
physical pion masses, thin lines to the chiral limit.}
\end{figure}
The change in the nucleon masses is
shown in Figure \ref{emn_vs_mu}.
The effective mass of the nucleon drops at the liquid gas phase 
transition, and does not change strongly as a function of 
the chemical potential until the chiral phase transition.
The mass of the parity partner also drops at the liquid gas phase 
transition, and then decreases strongly with increasing chemical 
potential.  In the chiral
limit, both nucleon masses jump discontinuously at the chiral transition,
to $m_0$.  If there is a smooth crossover, the mass of the nucleon 
and the one of its parity partners smoothly approach $m_0$ asymptotically,
the nucleon from below, and the parity partner, from above.
\begin{figure}[h]
\includegraphics[width=9cm]{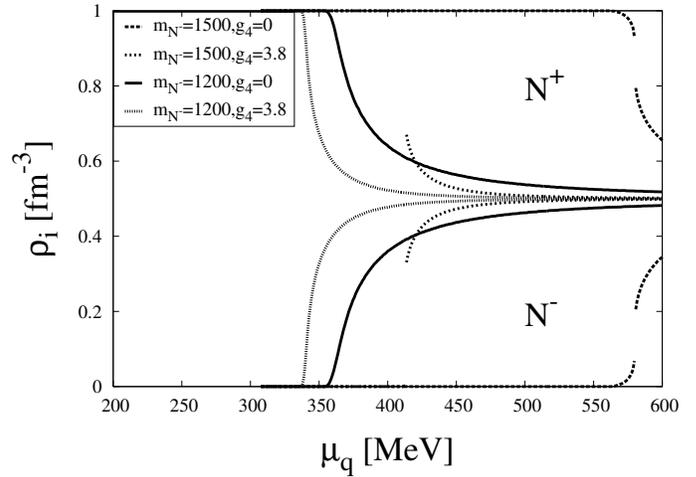}
\caption{\label{rhoi_vs_mu}
Relative densities of the 
nucleon and its parity partner, versus the quark chemical potential.}
\end{figure}

In Figure \ref{rhoi_vs_mu} we show the relative densities of the 
nucleon and its parity partner. From the figure, one can see that
the chiral phase transition occurs once there 
is any significant population of the $N^-$ states.
At asymptotically high densities or chemical potentials, the chiral condensate vanishes,
the nucleons are equal in mass, and so 
each chiral partner contributes half of the total nucleon density.
\begin{figure}[h]
\vspace{-1.0cm}
\includegraphics[width=9cm]{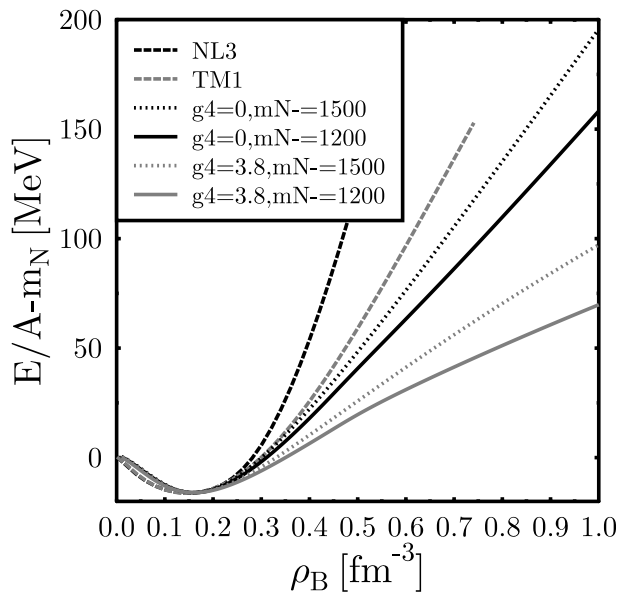}
\vspace{-0.6cm}
\caption{\label{eos_800}
Binding energy per particle of nuclear matter for $m_{N^-}=1200,1500$~MeV
and $g_4=0,3.8$  
in comparison with relativistic mean-field calculations 
TM1 \cite{TM1} and NL3 \cite{NL3}.}
\end{figure}

Figure \ref{eos_800} shows the resulting binding energy per particle 
for the parameter sets considered before.  
These are compared to the Walecka model
fits NL3 and TM1. At high densities 
all equations of state (EoS) in the parity doublet 
model are  much softer than the Walecka models, but  
they show a larger curvature at small densities, which causes 
the still relatively high incompressibilities.
As could be expected, for finite values of $g_4$, 
the EoS is considerably softened at high densities. 
A smaller mass of the negative parity state 
yields a reduced energy per particle at high densities.
\begin{figure}[h]
\includegraphics[width=9cm]{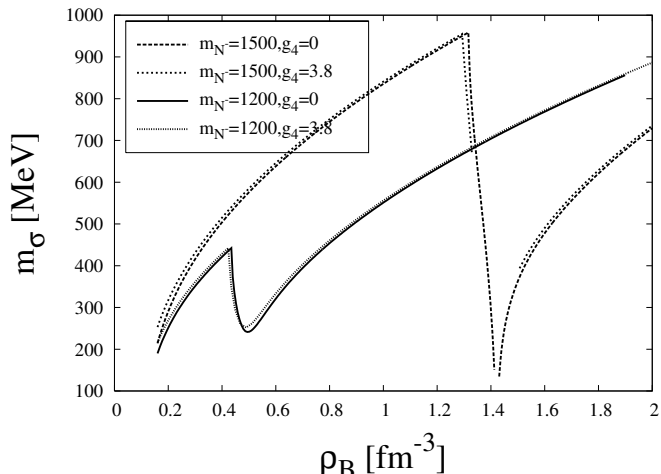}
\vspace{-0.6cm}
\caption{\label{msigma_vs_rho}
Effective sigma mass $m_\sigma^\ast$ vs density.
After a jump down at the liquid gas phase transition the 
mass increases again as a function of density until 
the $N^-$ states get populated. This causes a significant decrease
of $m_\sigma^\ast$. If the chiral phase transition actually takes place, the 
mass increases again.}
\end{figure}
Finally, we consider the 
behavior of the effective sigma mass $m_\sigma^\ast$ with density.
It is obtained by first determining the sigma-omega mass matrix
through the corresponding second derivatives of the 
thermodynamic potential (or pressure) with respect to the fields 
at fixed chemical potential and then 
diagonalizing this matrix.
In Figure \ref{msigma_vs_rho} we show the resulting 
effective $\sigma$-mass, $m^*_{\sigma}$ as a function density.
The different transitions in dense matter cause very 
significant structures. 
First, at the liquid gas phase transition, the
sigma mass jumps from its vacuum value down to a value of 
around 200-300 MeV
and then increases again. Right before the chiral transition,  
it decreases strongly.  
This takes place when the $N^-$ states start to get populated. 
This decrease continues until the chiral transition.
If the transition is of first order ($m_{N^-}=1.5$~GeV), 
the sigma mass jumps as well as the density and then starts to increase again. 
If the transition is a crossover ($m_{N^-}=1.2$~GeV), 
this increase happens continuously.
\section{Summary}
We have shown that it is possible to
obtain successful fits of saturated nuclear matter
in a SU(2) parity doublet model
with $\sigma$ and $\omega$ mesons. 
Agreement with current estimates of the nuclear 
incompressibility favors large values of the explicit mass 
term parameter $m_0$.  Furthermore, we found that 
nuclear matter fits are possible for different values 
of the vacuum mass of the N$^-$. At higher densities, 
chiral restoration takes place, where the order of 
the transition and the critical density depend 
on N$^-$'s mass.

There are clearly many avenues for further investigation: including
strange quarks, other hadronic resonances, and the like.  Probably
the outstanding question is to look at decay widths, given that
the natural experimental candidate for the nucleon's parity partner,
the N'(1535), likes to decay to $\eta \pi$ so much.  We simply found
the present exercise most encouraging, in that although the nucleon
parity partner is relatively heavy, the properties of nuclear matter
change significantly, and in a direction which bring them closer 
to known experimental values.
\section*{Acknowledgements}
D.Z acknowledges support from GSI and CNPq. 
L.T. was partially supported by the Alexander von Humboldt Foundation.
The research of R.D.P. was supported by the U.S. Department
of Energy grant DE-AC02-98CH10886; he also
thanks the Alexander von Humboldt Foundation for their support.

\end{document}